%% file: draft1.tex
\documentclass[conference]{IEEEtran}

\DeclareMathAlphabet{\mathitbf}{OML}{cmm}{b}{it}

\usepackage{tikz}
\usetikzlibrary{matrix,arrows}

\usepackage{amsmath,amssymb,mathrsfs,amsthm}
\usepackage[colorlinks=true,linkcolor=black,anchorcolor=black,citecolor=black,filecolor=black,menucolor=black,urlcolor=black]{hyperref}
\usepackage{graphicx}
\usepackage{psfrag}
\usepackage{subfigure}
\usepackage{cite}
\usepackage{multirow}
\usepackage{stfloats}
\usepackage[margin=1in]{geometry}
\usepackage{caption} 

\usepackage{epsfig}
\usepackage{amsfonts}
\usepackage{graphics}
\usepackage{color}
\usepackage{mathtools}
 
\DeclarePairedDelimiter{\floor}{\lfloor}{\rfloor} 
\epsfverbosetrue

\include{defns}  

\newcommand{\bcc}{b_\mathrm{CC}}

\newcommand{\Gu}{\mathrm{Un}(G)}

\def\0{\bf{0}}

\newtheorem{theorem}{Theorem}
\newtheorem{lemma}{Lemma}
\newtheorem{proposition}{Proposition}

\theoremstyle{definition}
\newtheorem{definition}{Definition}

\newtheorem{corollary}{Corollary}

\newtheorem{remark}{Remark}

\begin{document}

\title{Approximate Capacity of Index Coding \\ for Some Classes of Graphs}
\author{
\authorblockN{Fatemeh Arbabjolfaei and Young-Han Kim}
\authorblockA{Department of Electrical and Computer Engineering\\
University of California, San Diego\\
Email: \{farbabjo, yhk\}@ucsd.edu
}
}
\date{}
\maketitle

\begin{abstract}
For a class of graphs for which the Ramsey number $R(i,j)$ is upper bounded by $ci^aj^b$, for some constants $a,b,$ and $c$, it is shown that the clique covering scheme approximates the broadcast rate of every $n$-node index coding problem in the class within a multiplicative factor of $c^{\frac{1}{a+b+1}} n^{\frac{a+b}{a+b+1}}$ for every $n$. 
Using this theorem and some graph theoretic arguments, it is demonstrated that the broadcast rate of planar graphs, line graphs and fuzzy circular interval graphs is approximated by the clique covering scheme within a factor of $n^{\frac{2}{3}}$.
\end{abstract}

\section{Introduction}
Index coding is a broadcasting problem with side information
in which a server has $n$ messages $x_1, \ldots, x_n$, $x_j \in \{0,1\}^{t_j}$, that are to be sent to their respective receivers.
Receiver $j$ that is interested in message $x_j$ has side information about a subset $x(A_j)$ of other messages, $A_j \subseteq [n] \setminus \{j\}$.
The goal is to find the minimum number of bits that the server needs to broadcast to the receivers such that every receiver can recover its desired message using the received bits and its own side information.

\begin{figure}[h]
\begin{center}
\small
\psfrag{1}[cb]{1}
\psfrag{2}[rc]{2}
\psfrag{3}[lc]{3}
\psfrag{4}{4}
\includegraphics[scale=0.35]{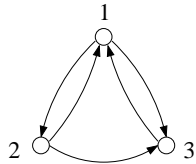}
\end{center}
\caption{The  graph representation for the index coding problem with $A_1 = \{2,3\}, A_2 = \{1\}$, and $A_3 = \{1,2\}$.}
\label{fig:3-message}
\end{figure}

Any instance of this problem, referred to collectively as the \emph{index coding problem}, 
is fully specified by the side information sets  $A_1, \ldots, A_n$. Equivalently, it can be specified
by a side information graph $G$ with $n$ nodes, in which
a directed edge $i \to j$ represents that receiver $j$ has message $i$ as side information, i.e., $i \in A_j$
(see Fig.~\ref{fig:3-message}).
Thus, we often identify an index coding problem with its side information graph and simply write ``index coding problem $G$.''

A $(t, r)$ index code is defined by
\begin{itemize}
\item an encoder $\phi: \{0,1\}^{tn} \to \{0,1\}^r$ that maps $n$-tuple of messages $x^n$
to an $r$-bit index and
\item $n$ decoders $\psi_j: \{0,1\}^r \times \{0,1\}^{t|A_j|} \to \{0,1\}^{t}$ that maps the received index $\phi(x^n)$ and the side information $x(A_j)$ back to $x_j$ for $j \in [n]$.
\end{itemize}
Thus, for every $x^n \in \{0,1\}^{tn}$,
\[
\psi_j(\phi(x^n), x(A_j)) = x_j, \quad j \in [n].
\]

The performance of an index code $\Cc$ is measured by its rate $\b(\Cc) = r/t$.
Define the \emph{broadcast rate} of the index coding problem as
\[
\b = \inf_t \inf \b(\Cc),
\]
where the second infimum is over all $(t,r)$ index codes.
Thus, $\b$ characterizes the fundamental limit on the rate of index codes such that every message can be recovered exactly.
($\frac{1}{\b}$ is also referred to as symmetric capacity.)

Despite numerous contributions made during the past two decades, no computable characterization of the broadcast rate or its approximation  within a factor of $O(n^{1-\epsilon})$, for some $\epsilon > 0$ exists.
The solution is only known in a handful of cases with special structures \cite{Arbabjolfaei--Bandemer--Kim--Sasoglu--Wang2013, Ong2014, Yi--Sun--Jafar--Gesbert2015, Maleki--Cadambe--Jafar2014,  Unal--Wagner2014, Ong--Lim--Ho2013}.
As an example, for the class of (undirected) perfect graphs the clique covering upper bound matches the maximum acyclic induced subgraph (MAIS) lower bound and thus the broadcast rate (and in general the capacity region) is known \cite{Yi--Sun--Jafar--Gesbert2015, Bar-Yossef--Birk--Jayram--Kol2011}. 

Blasiak, Kleinberg, and Lubetzky~\cite{Blasiak--Kleinberg--Lubetzky2013} proposed an algorithm that approximates the broadcast rate of a general index coding problem within a factor of $O\left(n\frac{\log \log n}{\log n}\right)$. 
For the class of undirected graphs, using Ramsey theory, they proposed a slightly better algorithm that approximates the broadcast rate within a factor of $O\left(\frac{n}{\log n}\right)$.
In this paper, we use the same technique to show that if the Ramsey number for a class of graphs  satisfies $R(i,j) \leq ci^aj^b$ for some constants $a,b$, and $c$, then  the clique covering scheme approximates the broadcast rate of every $n$-node graph in that class within a multiplicative factor of $c^{\frac{1}{a+b+1}} n^{\frac{a+b}{a+b+1}}$. 
From the Ramsey theory literature, we know that the classes of planar graphs, line graphs, and fuzzy circular interval graphs satisfy the condition and hence our result implies that for any index coding problem that lies in one of these classes, the clique covering scheme approximates the broadcast rate within a multiplicative factor of $n^{\frac{2}{3}}$.
Moreover, we show that for the class of planar graphs (complements of planar graphs), due to the four color theorem, uncoded transmission (clique covering) approximates the broadcast rate within the constant factor of four which is a much better approximation than the one derived using Ramsey theory.
We then generalize this result to directed graphs for which their corresponding undirected graph or the undirected graph corresponding to their complement are planar.
Finally, we establish a nontrivial lower bound on the broadcast rate of the unidirected graphs (graphs with no bidirectional edge) which implies that for the index coding problems in this class, the performance of uncoded transmission of the messages is within a multiplicative factor of $O\left(\frac{n}{\log n}\right)$ of the performance of the optimal coding scheme.

Throughout the paper, $G|_S$ denotes the subgraph of $G$ induced by the subset $S$ of vertices.

\section{Mathematical Preliminaries}

\subsection{Some Graph Classes}

\begin{figure}[h]
\vspace{0.75em}
\begin{center}
\subfigure[]{
\small
\psfrag{1}[bc]{1}
\psfrag{2}[bc]{2}
\psfrag{3}[c]{3}
\psfrag{4}[c]{4}
\includegraphics[scale=0.45]{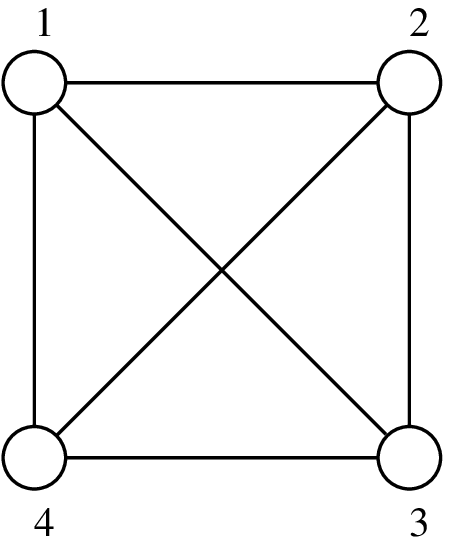}
}
\subfigure[]{
\hspace{2em}
\small
\psfrag{1}[bc]{1}
\psfrag{2}[bl]{2}
\psfrag{3}[tl]{3}
\psfrag{4}[tr]{4}
\psfrag{5}[br]{5}
\includegraphics[scale=0.45]{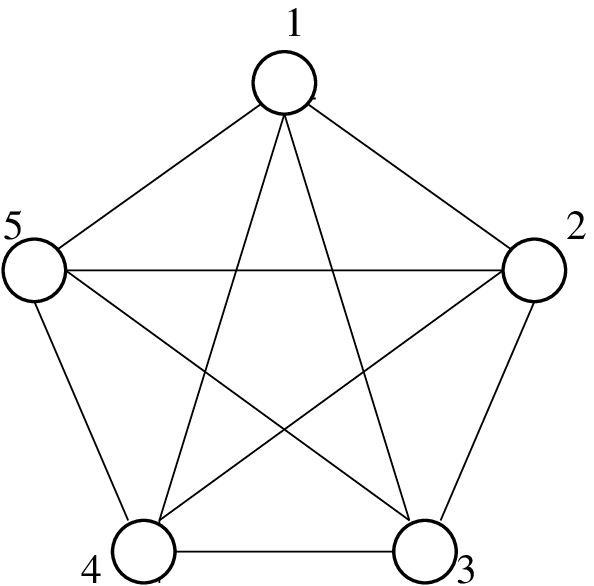}
}
\vspace{-1em}
\end{center}
\caption{(a) A 4-node planar graph (edge $\{1,3\}$ can be drawn such that it does not cross $\{2,4\}$). (b) A 5-node non-planar graph.}
\label{fig:4-node-planar}
\vspace{-1em}
\end{figure}

An undirected graph is said to be planar if it can be drawn in a plane without graph edges crossing, i.e., edges intersect only at the nodes.
Fig.~\ref{fig:4-node-planar} shows an example of a 4-node planar graph and a 5-node graph that is not planar.

\begin{figure}[h]
\vspace{0.75em}
\begin{center}
\subfigure[]{
\small
\psfrag{1}[bc]{1}
\psfrag{2}[bc]{2}
\psfrag{3}[c]{3}
\psfrag{4}[c]{4}
\psfrag{a}[bc]{a}
\psfrag{b}[cl]{b}
\psfrag{c}[cc]{c}
\psfrag{d}[cr]{d}
\psfrag{e}[br]{e}
\includegraphics[scale=0.45]{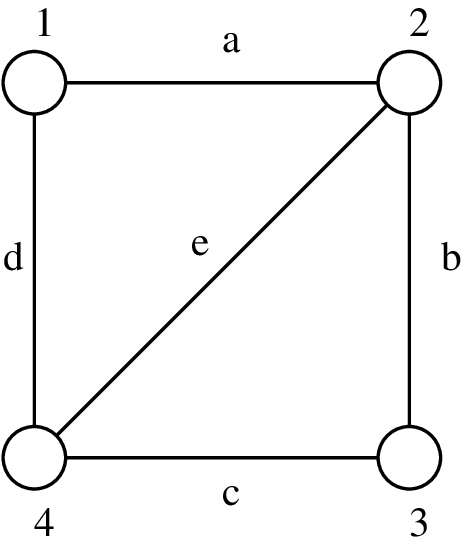}
}
\subfigure[]{
\hspace{3.5em}
\small
\psfrag{1}[bc]{1}
\psfrag{2}[bl]{2}
\psfrag{3}[tl]{3}
\psfrag{4}[tr]{4}
\psfrag{5}[br]{5}
\psfrag{a}[bc]{a}
\psfrag{b}[cl]{b}
\psfrag{c}[cc]{c}
\psfrag{d}[cr]{d}
\psfrag{e}[br]{e}
\includegraphics[scale=0.45]{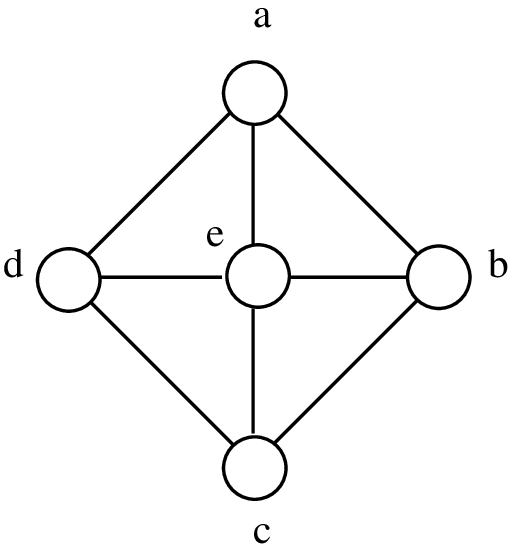}
}
\vspace{-1em}
\end{center}
\caption{(a) A 4-node graph with 5 edges. (b) The corresponding 5-node line graph.}
\label{fig:line}
\vspace{-1em}
\end{figure}

The line graph of an undirected graph $G$ is obtained by associating a vertex with each edge of the graph $G$ and connecting two vertices with an edge iff the corresponding edges of $G$ have a vertex in common. Fig.~\ref{fig:line} shows a graph and its corresponding line graph.

Given a circle $C$, a closed interval of $C$ is a proper subset of $C$ homeomorphic to the closed unit interval $[0,1]$; in particular, every closed interval of $C$ has two distinct endpoints. 
The class of fuzzy circular interval graphs introduced by Chudnovsky and Seymour \cite{Chudnovsky--Seymour2005} is a basic class of claw-free graphs and is defined as follows.

A graph $G=(V,E)$ is a fuzzy circular interval graph if the following conditions hold:
\begin{itemize}
\item there is a (not necessarily injective) mapping $\phi$ from $V$ to a circle $C$;
\item there is a set $F$ of closed intervals of $C$, none including another, such that no point of $C$ is an endpoint of more than one interval in $F$, and
\begin{itemize}
\item if two vertices $u, v \in V$ are adjacent, then $\phi(u)$ and $\phi(v)$ belong to a common interval of $F$;
\item if two vertices $u, v \in V$ are not adjacent, then either there is no interval in $F$ that contains both $\phi(u)$ and $\phi(v)$, or there is exactly one interval in $F$ whose endpoints are $\phi(u)$ and $\phi(v)$.
\end{itemize}
\end{itemize}

\begin{figure}[h]
\centering
\subfigure[]{
\small
\psfrag{1}[bc]{1}
\psfrag{2}[bc]{2}
\psfrag{3}[c]{3}
\psfrag{4}[c]{4}
\psfrag{5}[b]{5}
\psfrag{6}[b]{6}
\includegraphics[scale=0.45]{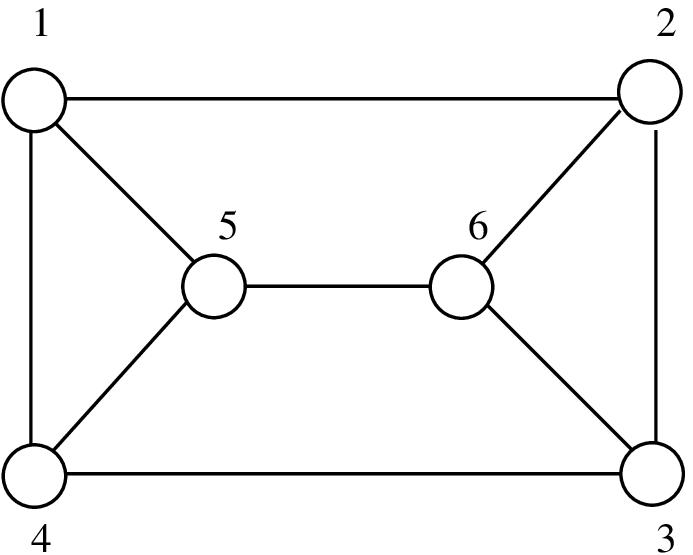}
}
\subfigure[]{
\small
\psfrag{x1}[tc]{$x_1$}
\psfrag{x2}[tc]{$x_2$}
\psfrag{x3}[tl]{$x_3$}
\psfrag{x4}[tl]{$x_4$}
\psfrag{x5}[bl]{$x_5$}
\psfrag{x6}[bc]{$x_6$}
\includegraphics[scale=0.40]{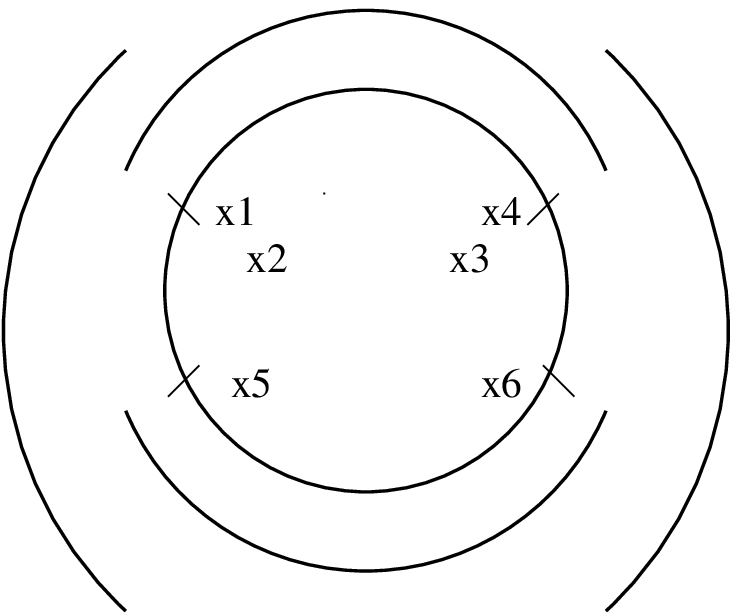}
}
\caption{(a) The complement of $C_6$. (b) The fuzzy circular interval model of $\bar{C_6}$.}
\label{fig:c6bar}
\end{figure}

See Fig. \ref{fig:c6bar} for a fuzzy circular interval model of the complement of $C_6$.
In this example, $V = [6]$ and $j$ is mapped to $x_j$ which is not an injective mapping.

\subsection{Graph Coloring}

A subset $I$ of the vertices of a graph $G=(V,E)$ is said to be independent if no two vertices of $I$ are adjacent.
The maximum size of an independent set of a graph $G$ is referred to as the independence number of the graph and is denoted by $\alpha(G)$.

A (vertex) coloring of an undirected graph $G$ 
is a mapping that assigns a color to each vertex
such that no two adjacent vertices share the same color.
The \emph{chromatic number} $\chi(G)$ is the minimum number of colors such that a coloring of the graph exists.

More generally, a $b$-fold coloring assigns a set of $b$ colors to each vertex 
such that no two adjacent vertices share the same color.
The $b$-fold chromatic number $\chi^{(b)}(G)$ is the minimum number of colors such that a $b$-fold coloring
exists.
The \emph{fractional chromatic number} of the graph is defined as
\[
\chi_f(G) = \lim_{b \rightarrow \infty} \frac{\chi^{(b)}(G)}{b} = \inf_b \frac{\chi^{(b)}(G)}{b},
\]
where the limit exists since $\chi^{(b)}(G)$ is subadditive.
Consequently,
\begin{equation} 
\label{eq:fractional}
\chi_f(G) \le \chi(G).
\end{equation}
Let $\Ic$ be the collection of all independent sets in $G$. 
The chromatic number and the fractional chromatic number 
are also characterized as the solution to the following optimization problem
\begin{equation*} 
\begin{split}
\text{minimize~}& \sum_{S \in \Ic} \rho_S\\
\text{subject to~} 
& \sum_{S \in \Ic \suchthat j \in S} \rho_S \ge 1, \quad j \in [1:n].
\end{split}
\end{equation*}
When the optimization variables $\rho_S$, $S \in \Ic$, take
integer values $\{0,1\}$, then the (integral) solution is the chromatic number. If this constraint is relaxed
and $\rho_S \in [0,1]$, then the (rational) solution is the fractional chromatic number \cite{Scheinerman--Ullman2011}.

\begin{lemma}[Scheinerman and Ullman \cite{Scheinerman--Ullman2011}]
\label{lem:fracchi}
For any graph $G$ with $n$ nodes, 
\begin{align*}
\chi_f(G) \geq \frac{n}{\alpha(G)}.
\end{align*}
\end{lemma}

The {\em four color theorem} states that the chromatic number of any planar graph is upper bounded by four.

\begin{lemma}[Appel, Haken, and Koch \cite{Appel--Haken--Koch1977}]
\label{lem:fourcolor}
Every planar graph $G$ is four-colorable, i.e., $\chi(G) \leq 4$.
\end{lemma}

\subsection{Ramsey Numbers}

Let $\mathcal{G}$ be a class of graphs, such as perfect graphs and line graphs;
see \cite{Belmonte--Heggernes--Hof--Rafiey--Saei2014} for an overview of graph classes.

\begin{definition}
For any graph class $\mathcal{G}$ and any two positive integers $i$ and $j$, the Ramsey number $R_\mathcal{G}(i, j)$ is the
smallest positive integer such that every graph in $\mathcal{G}$ on at least $R_\mathcal{G}(i, j)$ vertices has a clique
of size $i$ or an independent set of size $j$.
\end{definition}

If $\mathcal{G}$ is the class of all undirected, finite, and simple graphs, then the Ramsey  number is simply denoted by $R(i,j)$.
The following lemma presents an upper bound on the Ramsey numbers.

\begin{lemma}[Erdos and Szekeres~\cite{Erdos--Szekeres1935}]
\label{lem:ramseyupper}
For any $i,j$ we have 
\[
R(i,j) \leq \binom{i+j-2}{i-1}.
\]
\end{lemma}

The following lemma uses Ramsey numbers to indicate a relationship between the independence number of an undirected graph and the chromatic number of its complement.

\begin{lemma}[Alon and Kahale \cite{Alon--Kahale1998}]
\label{lem:tkm}
Let $t_k(m) = \max\{j \suchthat R(k,j) \leq m\}$.
If $\chi(\bar{G}) \geq n/k + m$, then an independent set of size $t_k(m)$ can be
found in $G$.
\end{lemma}

In general, determining Ramsey numbers is very hard and they are known only for very small values of $i$ and $j$ \cite{Belmonte--Heggernes--Hof--Rafiey--Saei2014}.
If either $i \leq 2$ or $j \leq 2$, then it is straight forward to calculate the Ramsey number $R(i,j)$.

\begin{remark}
\label{remark:1}
For any graph class $\mathcal{G}$
\begin{align*}
R_\mathcal{G}(1,j) = R_\mathcal{G}(i,1) = 1, \quad i,j \geq 1,
\end{align*}
if $\mathcal{G}$ contains all edgeless graphs, then
\begin{align*}
R_\mathcal{G}(2,j) = j, \quad j \geq 1,
\end{align*}
and if $\mathcal{G}$ contains all complete graphs, then
\begin{align*}
R_\mathcal{G}(i,2) = i, \quad i \geq 1.
\end{align*}
\end{remark}

Let $\mathcal{P}$ be the class of planar graphs.
As $\mathcal{P}$ contains all edgeless graphs, the Ramsey number for this class is completely determined by Remark \ref{remark:1} and the following.

\begin{lemma}[Steinberg and Tovey \cite{Steinberg--Tovey1993}]\ \\
\label{lem:planar}
\begin{itemize}
\item $R_\mathcal{P}(i, 2) = i, \quad i \leq 4, j \geq 1$,
\item $R_\mathcal{P}(3, j) = 3j-3, \quad j \geq 1$,
\item $R_\mathcal{P}(i, j) = 4j-3, \quad i \geq 4, j \geq 1,  (i,j) \not = (4,2)$.
\end{itemize}
\end{lemma}

Let $\mathcal{L}$ be the class of line graphs.
Since this class contains all edgeless graphs and all complete graphs, Remark \ref{remark:1} together with the following lemmas determine Ramsey numbers for this class for all pairs $(i,j)$.

\begin{lemma}[Matthews and Sumner \cite{Matthews--Sumner1985}] 
\label{lem:line1}
For every integer $j \geq 1$, $R_\mathcal{L}(3, j) = \floor{(5j - 3)/2}$.
\end{lemma}

\begin{lemma}[Belmonte, Heggernes, Hof, Rafiey, and Saei \cite{Belmonte--Heggernes--Hof--Rafiey--Saei2014}] 
\label{lem:line2}
For every pair of integers $i \geq 4$ and $j \geq 1$,
\begin{align*}
R_\mathcal{L}(i,j) = \begin{cases}
i(j - 1) - (t + r) + 2 & \text{if}~ i = 2k, \\
i(j - 1) - r + 2 & \text{if}~ i = 2k + 1,
\end{cases}
\end{align*}
where 
$j = tk + r$, $t \geq 0$ and $1 \leq r \leq k$.
\end{lemma}

Let $\mathcal{F}$ be the class of fuzzy circular interval graphs. 
This class contains all edgeless graphs and all complete graphs.
Hence,  Remark \ref{remark:1} and the following lemma determine Ramsey numbers for the graphs in this class.

\begin{lemma}[Belmonte, Heggernes, Hof, Rafiey, and Saei \cite{Belmonte--Heggernes--Hof--Rafiey--Saei2014}]
\label{lem:fuzzy-circular-ramsey}
For every pair of integers $i, j \geq 3$,
\begin{align*}
R_\mathcal{F}(i,j) = (i-1)j.
\end{align*}
\end{lemma}

Therefore, for every pair of integers $(i,j)$, $R_\mathcal{P}(i,j)$, $R_\mathcal{L}(i,j)$ and $R_\mathcal{F}(i,j)$ are determined by Lemmas \ref{lem:planar} through \ref{lem:fuzzy-circular-ramsey} and Remark \ref{remark:1}. 
Hence we can derive a simple bilinear upper bound on the Ramsey numbers of any member of these classes.
 
\begin{lemma}
\label{lem:bilinearupper}
For every pair of integers $(i,j)$,  $i,j \geq 1$,
and for $\Gc = \Pc, \Lc$, or $\Fc$,
\begin{align*}
R_\Gc(i,j) &\leq ij.
\end{align*}
\end{lemma}

\section{Existing Bounds on the Broadcast Rate}

The simplest approach to index coding is a coding scheme by Birk and Kol \cite{Birk--Kol1998} that partitions the side information
graph $G$ by cliques
and transmit the binary sums (parities)
of all the messages in each clique.

\begin{proposition}[Clique covering bound]
\label{prop:cc}
Let $\bcc(G)$ be the minimum number of cliques that partition $G$ 
(or equivalently, the chromatic number of the undirected complement of $G$) which is the solution to the integer program
\begin{equation} \label{eq:chromatic}
\begin{split}
\text{minimize~~}& \sum_{S \in \Kc} \rho_S\\
\text{subject to~~} 
& \sum_{S \in \Kc \suchthat j \in S} \rho_S \ge 1, \quad j \in [n],\\
& \rho_S \in \{0,1\}, \quad S \in \Kc,
\end{split}
\end{equation}
where $\Kc$ is the collection of all cliques in $G$.
Then for any index coding problem $G$, $\b(G) \leq \bcc(G)$.
\end{proposition}

Blasiak, Kleinberg, and Lubetzky~\cite{Blasiak--Kleinberg--Lubetzky2013} extended this bound to the \emph{fractional clique covering bound} (which is equivalent to the fractional chromatic number of the undirected complement of $G$) by 
relaxing the integer constraint $\rho_S \in \{0,1\}$ in \eqref{eq:chromatic} to $\rho_S \in [0,1]$.
In \cite{Bar-Yossef--Birk--Jayram--Kol2011}, Bar-Yossef, Birk, Jayram, and Kol proposed the following lower bound on the broadcast rate of the index coding.

\begin{proposition}[Maximum acyclic induced subgraph (MAIS) bound]
\label{prop:mais}
For any index coding problem $G$
\[
\max_{S \subseteq [1:n] \suchthat G|_S ~\text{is cycle-free} } |S| \leq \b.
\]
\end{proposition} 

\section{Undirected Graphs}

If every edge of the side information graph is bidirectional, then we consider the graph as an undirected graph.
We begin with the statement of an approximation result for the class of undirected graphs
and its proof, as the same technique will be used to generate the main result of this paper.

\begin{proposition}[Blasiak, Kleinberg, and Lubetzky~\cite{Blasiak--Kleinberg--Lubetzky2013}]
\label{prop:undir}
For any undirected graph $G$ with $n$ nodes, the clique covering scheme approximates the broadcast rate of the index coding problem $G$ within a factor of 
$O\left(\frac{n}{\log n}\right)$.
\end{proposition}

\begin{IEEEproof}
Combine Lemma \ref{lem:ramseyupper} and Lemma \ref{lem:tkm}  and take $m=n/k$ with $k = \frac{1}{2} \log n$. Then either $\chi(\bar{G}) < \frac{4n}{\log n}$ or there exists an independent set of size $t_{k}(n/k) \geq \max\{j \suchthat \binom{0.5 \log n + j -2}{0.5 \log n -1} \leq \frac{2n}{\log n}\} \geq 0.5 \log n$  for sufficiently large $n$.
In both cases $\frac{\chi(\bar{G})}{\b} \leq O\left(\frac{n}{\log n}\right),$ which completes the proof of the proposition.
\end{IEEEproof}


Next, we present conditions under which there exists an approximation of the broadcast rate within a factor of $O(n^{1-\epsilon})$ for some $\epsilon > 0$.

\begin{theorem}
\label{thm:main}
Let $\mathcal{G}$ be a class of graphs for which $R(i,j) \leq ci^aj^b$ holds for some constants $a,b$, and $c$. 
Then the clique covering scheme approximates the broadcast rate of every $n$-node problem in $\mathcal{G}$ within a multiplicative factor of $c^{\frac{1}{a+b+1}} n^{\frac{a+b}{a+b+1}}$.
\end{theorem}

\begin{IEEEproof}
Let $G \in \mathcal{G}$.
If $\alpha(G) = 1$, then the graph is a clique and $\b = 1$.
Thus we assume that $\alpha \geq 2$.
Let $k$ be a positive real number.
Consider two cases. \\
\em{Case 1:} If $\chi(\bar{G}) < 2n/k$, then we have
\begin{align*}
\label{eq:case1}
2 \leq \alpha(G) \leq \b \leq \chi(\bar{G}) < 2n/k.
\end{align*}
\em{Case 2:} If $\chi(\bar{G}) \geq 2n/k$, then by letting $m = n/k$ in Lemma \ref{lem:tkm} we have $t_k(n/k) \leq \alpha(G)$.
Hence
\begin{align}
\left( \frac{n}{ck^{a+1}} \right)^{\frac{1}{b}}
&= \max \{j \suchthat ck^aj^b \leq n/k\} \nonumber \\
&\leq \max \{j \suchthat R(k,j) \leq n/k\} \nonumber \\
&=  t_k(n/k) 
\leq \alpha(G) 
\leq \b
\leq n. \nonumber
\end{align}

To minimize the multiplicative gap in both cases we equate the ratio of the upper bound to the lower bound in the two cases, which gives
\[
k = \left(\frac{n}{c}\right)^{\frac{1}{a+b+1}},
\] 
and thus yields the multiplicative gap of $c^{\frac{1}{a+b+1}} n^{\frac{a+b}{a+b+1}}$ and completes the proof of the theorem. 
\end{IEEEproof}

As stated in Lemma \ref{lem:bilinearupper}, planar graphs, line graphs and fuzzy circular interval graphs are three classes that satisfy the condition of Theorem \ref{thm:main} with $a=b=c=1$.

\begin{corollary}
If $G$ is a planar graph or a line graph or a fuzzy circular interval graph with $n$ nodes, the clique covering scheme approximates the broadcast rate within a multiplicative factor of $n^{2/3}$.
\end{corollary}

However, if either the graph or its complement is planar, we can establish a stronger approximation result 
using the four color theorem (Lemma \ref{lem:fourcolor}).

\begin{theorem}
\label{thm:planar}
If either $G$ or its complement is planar, the broadcast rate can be approximated within a multiplicative factor of four.
\end{theorem}
\vspace{-0.25cm}
\begin{IEEEproof}
If $G$ is planar, we have
\begin{align*}
\frac{n}{4} \leq  \frac{n}{\chi(G)}
\leq \frac{n}{\chi_f(G)}
\leq \alpha(G) 
\leq \b 
\leq n,
\end{align*}
where the first inequality follows from Lemma \ref{lem:fourcolor}, the second inequality follows from \eqref{eq:fractional}, and the third one follows from Lemma \ref{lem:fracchi}.
If $\bar{G}$ is planar, we have
\[
1 \leq \alpha(G) \leq \b \leq \chi(\bar{G}) \leq 4.
\]
\end{IEEEproof}

Due to Theorem \ref{thm:planar}, if $G$ ($\bar{G}$) is planar then uncoded transmission (clique covering) is within a multiplicative factor of four from optimal. 
\vspace{-0.15cm}
\section{Directed Graphs}

In \cite{Blasiak--Kleinberg--Lubetzky2013}, Blasiak, Kleinberg, and Lubetzky generalized the result of Proposition \ref{prop:undir} to directed graphs.

\begin{proposition}[Blasiak, Kleinberg, and Lubetzky~\cite{Blasiak--Kleinberg--Lubetzky2013}]
For any index coding problem with $n$ messages, the fractional clique covering scheme approximates the broadcast rate within a multiplicative factor of $O\left(n\frac{\log \log n}{\log n} \right)$.
\end{proposition}


To the best of our knowledge, the above approximation is the only algorithm to approximate the broadcast rate of a general index coding problem.
In particular, no $O\left(n^{1-\epsilon} \right)$ approximation  exists for any $\epsilon > 0$.

Next, we generalize some of the results of the previous section to directed graphs.
Let $G$ be a directed side information graph.
Then by $\Gu$, we denote the undirected graph resulted from replacing all (bidirectional and unidirectional) edges of $G$ with an undirected edge.

\begin{theorem}
If either $\Gu$ or $\mathrm{Un}(\bar{G})$ is planar, the broadcast rate can be approximated within a multiplicative factor of four.
\end{theorem}

\begin{IEEEproof}
If $\Gu$ is planar,
\begin{align}
n/4 \leq \b(\Gu) 
\leq \b(G) 
\leq n \nonumber,
\end{align}
where the first inequality follows from Theorem \ref{thm:planar} and the second one holds since adding side information decreases the broadcast rate.
If $\mathrm{Un}(\bar{G})$ is planar,
\[
1 \leq \b(G) \leq \chi(\bar{G}) \leq \chi(\mathrm{Un}(\bar{G})) \leq 4.
\]
\end{IEEEproof}

%
%
\vspace{-0.1cm}
A graph $G=(V,E)$ is said to be unidirected if for any $i,j \in [n]$ such that $(i,j) \in E$, we have $(j,i) \not \in E$.
\begin{lemma}
\label{lem:unidir}
For any unidirected graph $G$ on $n=2^k$ nodes for some $k \in \mathbb{N}$,
$\log n \leq \beta$.
\end{lemma}

\begin{IEEEproof}
We use induction to prove that every unidirected graph with $n=2^k$ nodes contains an acyclic subgraph of size $\log n$ and then the lemma follows from the MAIS bound. 
This obviously holds for $n=2$.
Assume that it holds for $n = 2^{k-1}$.
Without loss of generality, we can assume that the underlying undirected graph is complete. 
Since if it is not complete we can always add edges without increasing the broadcast rate.
Pick an arbitrary node $v$ of the $n$-node graph. 
Due to the pigeonhole principle, this node has at least $\frac{n}{2}$ ingoing or outgoing edges.
Therefore, due to the induction hypothesis, there are $\log\left(\frac{n}{2} \right)$ nodes that form an acyclic vertex induced subgraph.
As all the edges connecting node $v$ to the induced subgraph are in one direction, we can add $v$ to the set of $\log\left(\frac{n}{2} \right)$ nodes and still have an acyclic vertex induced subgraph. 
Hence we have found an acyclic vertex induced subgraph of size $\log n$.
\end{IEEEproof} 

\begin{theorem}
\label{thm:unidir}
For any unidirected graph $G$ on $n$ nodes
$\floor{\log n} \leq \beta$.
\end{theorem}
\vspace{-0.2cm}
\begin{IEEEproof}
Assume that $\floor{\log n} = k$ and let $G'$ be an arbitrary subgraph of $G$ with $2^k$ nodes. Then
\begin{align*}
\floor{\log n} \leq \b(G') \leq  \b(G),
\end{align*}
where the first inequality follows from Lemma \ref{lem:unidir}.
\end{IEEEproof}

%

\begin{corollary}
For unidirected graphs, uncoded transmission is within a factor of $O\left(\frac{n}{\log n}\right)$ from optimal.
\end{corollary}

\newcommand{\BIBdecl}{\addtolength{\itemsep}{0.3mm}} 
\bibliographystyle{IEEEtran}
\bibliography{nit} 

\end{document}

%% file: defns.tex
\usepackage{xspace}
\usepackage{bbm}
\input{mathlig}

\newcommand{\muspace}{\mspace{1mu}}

\DeclareRobustCommand{\scond}{\mathchoice{\muspace\vert\muspace}{\vert}{\vert}{\vert}}
\mathlig{|}{\scond}

\DeclareRobustCommand{\discint}{\mathchoice{\mspace{-1.5mu}:\mspace{-1.5mu}}{\mspace{-1.5mu}:\mspace{-1.5mu}}{:}{:}}
\mathlig{::}{\discint}
\newcommand{\suchthat}{\mathchoice{\colon}{\colon}{:\mspace{1mu}}{:}}

\newcommand{\Cc}{\mathcal{C}}

\newcommand{\Fc}{\mathcal{F}}
\newcommand{\Gc}{\mathcal{G}}

\newcommand{\Ic}{\mathcal{I}}

\newcommand{\Kc}{\mathcal{K}}
\newcommand{\Lc}{\mathcal{L}}

\newcommand{\Pc}{\mathcal{P}}

\def\b{\beta}

\def\textiid{i.i.d.\@\xspace}
\newcommand\iid{\ifmmode\text{ i.i.d. } \else \textiid \fi}

\def\mathllap{\mathpalette\mathllapinternal}
\def\mathllapinternal#1#2{%
  \llap{$\mathsurround=0pt#1{#2}$}}

\def\clap#1{\hbox to 0pt{\hss#1\hss}}
\def\mathclap{\mathpalette\mathclapinternal}
\def\mathclapinternal#1#2{%
  \clap{$\mathsurround=0pt#1{#2}$}}

\let\oldstackrel\stackrel
\renewcommand{\stackrel}[2]{\oldstackrel{\mathclap{#1}}{#2}}

\renewcommand{\hbar}{h\mathllap{\overline{\vphantom{h}\hphantom{\rule{4.6pt}{0pt}}}\mspace{0.77mu}}}

\catcode`~=11 
\newcommand{\urltilde}{\kern -.06em\lower -.06em\hbox{~}\kern .02em}
\catcode`~=13 

%% file: mathlig.tex
\catcode`@=11
\chardef\mathlig@atcode\count255

\def\actively#1#2{\begingroup\uccode`\~=`#2\relax\uppercase{\endgroup#1~}}
\def\mathlig@gobble{\afterassignment\mathlig@next@cmd\let\mathlig@next= }
\def\mathlig@delim{\mathlig@delim}
\def\mathlig@defcs#1{\expandafter\def\csname#1\endcsname}
\def\mathlig@let@cs#1#2{\expandafter\let\expandafter#1\csname#2\endcsname}
\def\mathlig@appendcs#1#2{\expandafter\edef\csname#1\endcsname{\csname#1\endcsname#2}}

\def\mathlig#1#2{\mathlig@checklig#1\mathlig@end\mathlig@defcs{mathlig@back@#1}{#2}\ignorespaces}

\def\mathlig@checklig#1#2\mathlig@end{%
 \expandafter\ifx\csname mathlig@forw@#1\endcsname\relax
 \expandafter\mathchardef\csname mathlig@back@#1\endcsname=\mathcode`#1%
 \mathcode`#1"8000\actively\def#1{\csname mathlig@look@#1\endcsname}%
 \mathlig@dolig#1\mathlig@delim
\fi
\mathlig@checksuffix#1#2\mathlig@end
}

\def\mathlig@checksuffix#1#2\mathlig@end{%
\ifx\mathlig@delim#2\mathlig@delim\relax\else\mathlig@checksuffix@{#1}#2\mathlig@end\fi
}
\def\mathlig@checksuffix@#1#2#3\mathlig@end{%
\expandafter\ifx\csname mathlig@forw@#1#2\endcsname\relax\mathlig@dosuffix{#1}{#2}\fi
\mathlig@checksuffix{#1#2}#3\mathlig@end
}

\def\mathlig@dosuffix#1#2{%
\mathlig@appendcs{mathlig@toks@#1}{#2}%
\mathlig@dolig{#1}{#2}\mathlig@delim
}

\def\mathlig@dolig#1#2\mathlig@delim{%
 \mathlig@defcs{mathlig@look@#1#2}{%
 \mathlig@let@cs\mathlig@next{mathlig@forw@#1#2}\futurelet\mathlig@next@tok\mathlig@next}%
 \mathlig@defcs{mathlig@forw@#1#2}{%
  \mathlig@let@cs\mathlig@next{mathlig@back@#1#2}%
  \mathlig@let@cs\checker{mathlig@chck@#1#2}%
  \mathlig@let@cs\mathligtoks{mathlig@toks@#1#2}%
  \expandafter\ifx\expandafter\mathlig@delim\mathligtoks\mathlig@delim\relax\else
  \expandafter\checker\mathligtoks\mathlig@delim\fi
  \mathlig@next
 }%
 \mathlig@defcs{mathlig@toks@#1#2}{}%
 \mathlig@defcs{mathlig@chck@#1#2}##1##2\mathlig@delim{%
  \ifx\mathlig@next@tok##1%
   \mathlig@let@cs\mathlig@next@cmd{mathlig@look@#1#2##1}\let\mathlig@next\mathlig@gobble
  \fi 
  \ifx\mathlig@delim##2\mathlig@delim\relax\else
   \csname mathlig@chck@#1#2\endcsname##2\mathlig@delim
  \fi
 }%
 \ifx\mathlig@delim#2\mathlig@delim\else
  \mathlig@defcs{mathlig@back@#1#2}{\csname mathlig@back@#1\endcsname #2}%
 \fi
}%

\catcode`@\mathlig@atcode